\documentclass[a4paper]{spie}  

\usepackage{amsmath,amsfonts,amssymb}
\usepackage{graphicx}
\usepackage[colorlinks=true, allcolors=blue]{hyperref}

\title{New methods for ALMA angular-scale based observation scheduling, quality assessment, and beam shaping II: refinements}

\author[a]{Dirk Petry}
\author[a]{Mar\'ia D\'iaz Trigo}
\author[b,c]{R\"{u}diger Kneissl}
\author[c]{Ignacio Toledo}
\author[d]{Atsushi Miyazaki}
\author[d]{Toshinobu Takagi}
\author[a]{Ashley Barnes}
\author[e]{Francesca Bonanomi}
\affil[a]{European Southern Observatory, Karl-Schwarzschild-Str. 2, 85748 Garching, Germany}
\affil[b]{European Southern Observatory, Alonso de Cordova 3107, Vitacura, Santiago, Chile}
\affil[c]{Joint ALMA Observatory, Alonso de Cordova 3107, Vitacura, Santiago, Chile}
\affil[d]{Japan Space Forum, 3-2-1 Kanda-Surugadai, Chiyoda-ku, Tokyo 101-0062, Japan}
\affil[e]{Dept. of Astrophysics, University of Vienna, T\"{u}rkenschanzstr. 17, 1180, Vienna, Austria}

\authorinfo{Further author information: (Send correspondence to Dirk Petry)\\D. Petry: E-mail: dpetry@eso.org}

\begin{document} 
\maketitle

\begin{abstract}
The Atacama Large Millimeter/submillimeter Array remains the largest mm radio interferometer observatory
world-wide. It is now conducting its 11th observing cycle. 
In our previous paper presented at this conference series in 2020, we outlined a number of possible improvements to the ALMA end-to-end observing and data
processing procedures which could further optimize the uv coverage and thus the image quality while
at the same time improving the observing efficiency. Here we report an update of our results
refining our proposed adjustments to the scheduling and quality assurance processes. In particular
we present new results on ways to assess the uv coverage of a given observation efficiently, methods to define and measure the maximum recoverable angular scale, and on
the robustness of the deconvolution in the final interferometric imaging process w.r.t. defects
in the uv coverage. Finally we present the outline of a design for integrating uv coverage assessment into the control and processing loop of observation scheduling.
The results are applicable to all radio interferometers with more than approx. 10 antennas. 
\end{abstract}


\section{INTRODUCTION}
\label{sec:intro}

The Atacama Large Millimeter/submillimeter Array (ALMA) is an international collaboration between East Asia, Europe, and North America in cooperation with the Republic of Chile. The official project website for scientists is the {\it ALMA Science Portal}
{\tt http://www.almascience.org}. A detailed description of ALMA and the standard terms and procedures mentioned here can be found in the Cycle 11 ALMA Technical Handbook\cite{cortes2024} (THB).
A brief introduction to the project and the issues we are addressing with this work,
can be found in our first paper \cite{petry2020}.
In order to avoid repetition, we assume in the following that the reader is familiar with
ALMA observations and operations in general as they are described in the two references given above.

In this paper we present the results of an ESO ALMA internal development study on improving
ALMA's methods of assuring the achievement of the science goals of its users both in observation scheduling
and in subsequent quality assurance process (QA).

\section{The Baseline Length Distribution as a tool in scheduling and quality assurance}
\begin{figure} [ht]
   \begin{center}
   \begin{tabular}{c} 
   \includegraphics[height=5cm]{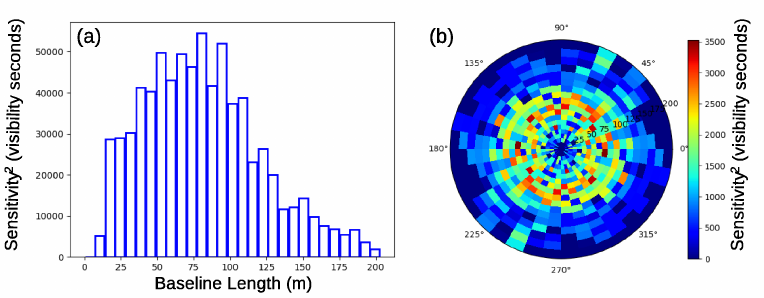}
   \end{tabular}
   \end{center}
   \caption[] { \label{fig:bldistexamples} 
   Example of a 1D Baseline Length Distribution (BLD) (a) and a 2D BLD (b), in this case for an 8 minute observation with the  ALMA 12M array in a compact configuration.
   }   
   \end{figure}

In order to assess the sensitivity of an observation at all angular scales, we introduce the {\it Baseline Length
Distribution} (BLD) in 1D (ignores baseline orientation) and 2D. The {\it observed BLD} of an ALMA
observation (Member Observation Unit Set, MOUS, composed of one or more Execution Blocks, EBs\cite{cortes2024}) is the histogram of the baseline lengths (BLs)  of the recorded interferometric visibilities for a representative channel in the representative spectral window and for a representative target 
{\it after removing invalid data} ("flagging"). Each entry is weighted by
its integration time multiplied by the relative visibility weight (weight\cite{casa2022} divided by the average weight for all baselines of the EB) and optionally by the squared ratio $(T_{sys,exp}/T_{sys,EB})^2$ of the expected system temperature $T_{sys,exp}$ and the actual average system temperature of all the antennas used in the EB.
The unit of the histogram content is thus "visibility seconds" or "visibility hours" (see, e.g., Fig. \ref{fig:bldistexamples}).

The 1D BLD permits more clear visualisation, precise reading of values, and overplotting of different BLDs. 
However, only the 2D BLD contains all information for a complete uv coverage assessment, and we present here an assessment tool for full 2D treatment.

The principle investigator (PI) of an ALMA observation does not propose for a fixed observing time
but defines "science goal" parameters for each MOUS which the observatory then aims to obtain in the most efficient way.  
An {\it expected BLD} can be derived from the science goal parameters Angular Resolution (AR), 
Largest Angular Scale (LAS), and sensitivity (translated internally into observing time)
for the representative observation target.
This BLD expectation can either optimise the Point Spread Function (PSF) shape or try to follow the typical
BLD shape produced by the actual observatory, which uses a compromise between optimal PSF
and being sensitive over a large range of angular scales.

Comparison of the observed and expected BLDs yields information about the BL range(s) where
the MOUS lacks sensitivity and where it is overexposed. To quantify the differences, we introduce the 
{\it Filling Fraction} (FF) as the ratio of observed and expected visibility seconds in an individual BL bin. If the expectation is zero, the FF is defined to be unity to indicate that in the given bin no sensitivity
is missing.

The weighting by $(T_{sys,exp}/T_{sys,EB})^2$ (mentioned above) needs to be applied
to take into account sensitivity differences between individual EBs and between the actual 
and typical observing conditions. The FF for the entire uv plane (all BLs) then becomes equivalent to the "execution fraction" (EF) which ALMA has been using
successfully since many Cycles to monitor the accumulated sensitivity and compare it to
the planned total sensitivity of an observation. 

Since we find that we need to separately assess PSF quality and fulfillment of scheduling goals,
we use the term FF referring to the filling fractions derived from the expectation for optimal PSF,
and the term EF referring to the filling fractions derived from the expectation for typical
observatory BLD shapes.

  \begin{figure} [ht]
   \begin{center}
   \begin{tabular}{cc} 
   \includegraphics[height=5cm]{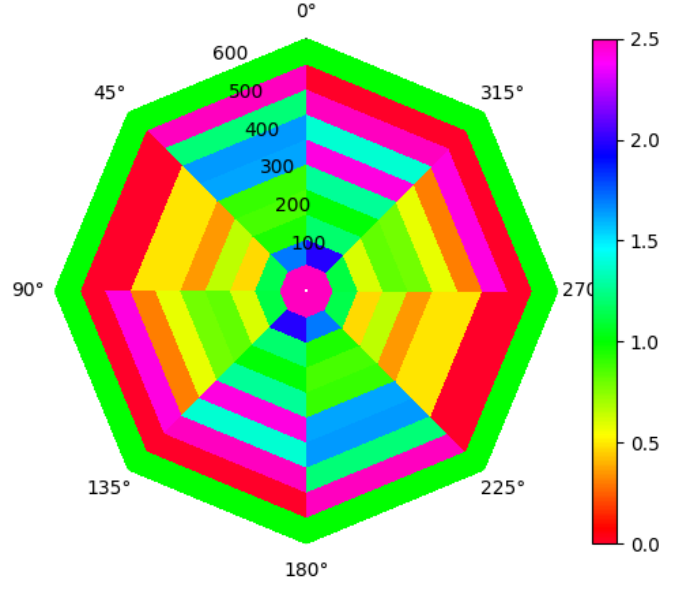} & \includegraphics[height=5cm]{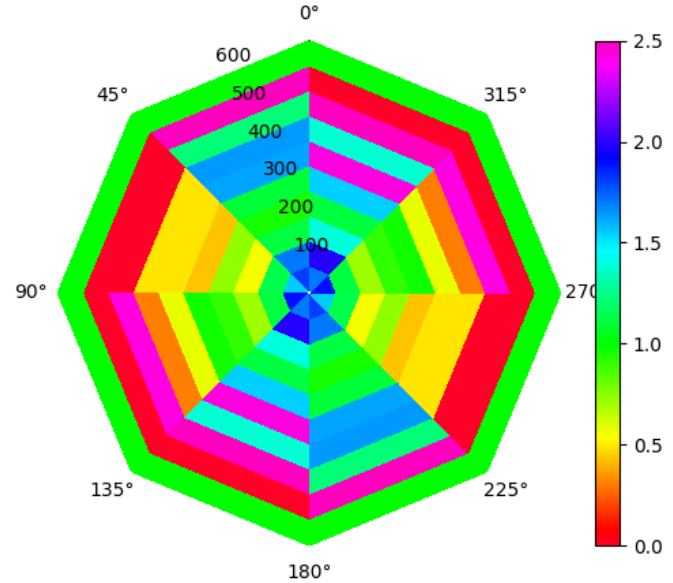}\\
   \end{tabular}
   \end{center}
   \caption[] { \label{fig:ffandefdistexamples} 
   Example of 2D Filling Fraction (left) and Execution Fraction (right) plots for a case with 
   sub-optimal uv coverage (a 24 minute observation with the ALMA 12M array with only 37 instead of normally at least 43 antennas in an intermediate configuration). There is underexposure (red, yellow) at the longest baselines in all sectors, and at intermediate baselines in one sector.
   }   
   \end{figure} 

Computing the EF in separate regions of the uv plane permits to generalise the ALMA concept of the 
single-number EF, which was used in scheduling so far, and replace it by a matrix of values, the EF~matrix. We recommend a 4~x~10 EF~matrix in four azimuthal bins of 45$^\circ$ width and 10 equidistant BL bins between 
the shortest possible BL and the longest BL, where the latter is the maximum BL of the expectation or observation, whichever is larger.

A quality assessment can then be performed by defining conditions on the FFs and EFs. 
If the conditions are not fulfilled for a given MOUS, one can derive the array configuration and hour-angle range in which additional EBs should be scheduled to obtain the missing visibility seconds most efficiently. For 
multi-EB MOUSs, this procedure can already be applied in the first level of QA (QA0) to steer the scheduling of subsequent EBs.

\subsection{Azimuthal uv coverage}

It is important to note that achieving azimuthal completeness of the uv coverage and achieving the overall sensitivity in minimal time are two competing goals because optimal azimuthal coverage requires to make use
of Earth rotation synthesis, even with the large
number of antennas available to ALMA. For image quality, complete azimuthal coverage, i.e. having sufficient sensitivity for a given angular scale at all baseline {\it orientations}, is key. Achieving azimuthal completeness should therefore take precedence over minimizing the total observation time.

\subsection{Maximum recoverable angular scale}

The second pair of (in general) competing conditions is that of the achievement of the most Gaussian PSF for a given AR and achieving an arbitrary maximum recoverable angular scale (MRS). With the best PSF, the MRS of an ALMA 12M array with a given AR
cannot be made much larger than ca. 8$\times$AR. To achieve larger values, the PSF has to be compromised and this
is what was done in the design of the ALMA 43-antenna configurations (C43s).
As mentioned above, we therefore need to distinguish between an expected BLD for best PSF 
and an expected BLD for the ALMA C43s. 

\subsection{Handling of mosaics}
In a mosaic observation, each field (pointing) constitutes in principle
a separate observation with its own uv coverage. However, since the time per visit to
each field is chosen to be short compared to the total observing time spent on the
mosaic, the difference which Earth rotation causes between the uv coverage patterns achieved for each individual field should be small.

Our new uv coverage assessment procedure includes the generation of a diagnostic plot to confirm this: the BLD for each mosaic field is measured separately and all BLDs are put in one plot such that differences between 
the uv coverage of the individual pointings
are immediately obvious. We find that for the ALMA 12M array, the deviations between the individual
BLDs of the mosaic fields are in general smaller than 1\% such that an assessment of the average BLD (over all fields) or simply the sum of the BLDs over all fields of a mosaic is a valid test of the uv coverage quality.

\section{Impact of BLD defects on the imaging results}
\label{defectimpact}
In order to decide on the tolerances within which the expected BLD is
to be achieved by an ALMA observation, we have studied the consequences
of non-ideal BLDs on the image quality and the derived scientific results.

  \begin{figure} [b]
   \begin{center}
   \includegraphics[height=9cm]{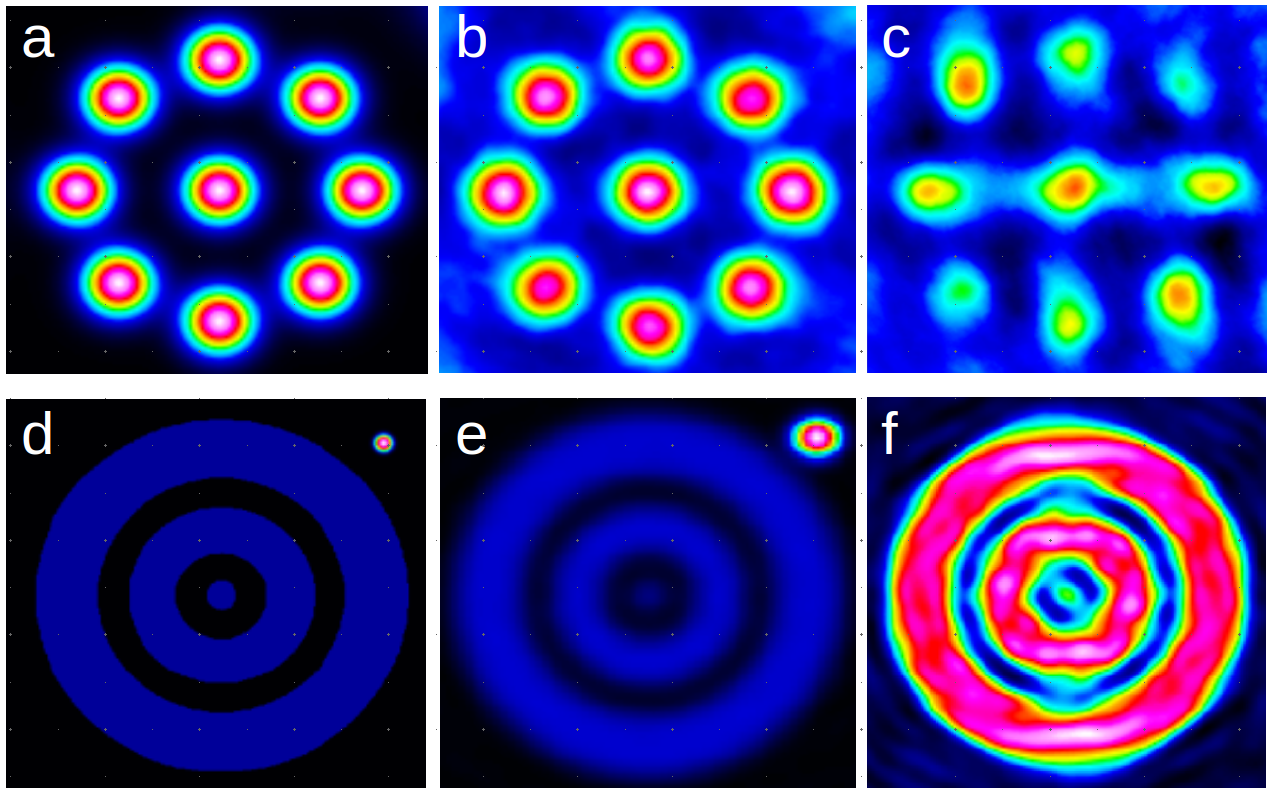} 
   \end{center}
   \caption[] { \label{fig:defectimpact} 
   Illustrations of the two different sub-studies performed to measure the impact
   of BLD defects on image quality via simulations: {\bf(a)} synthetic pattern of Gaussians used as input
   for an ALMA observation simulation, {\bf(b)} result of the simulated observation of (a)
   with a BLD of near-ideal shape, {\bf(c)} like (b) but for a non-ideal BLD, {\bf(d)} synthetic pattern to be used as
    input for an ALMA observation simulation resembling
   a protoplanetary disk with a nearby bright compact source, {\bf (e)} result of the simulated
   observation of (d) for an ideally shaped BLD, {\bf (f)} similar to (e) but with a non-ideal
   BLD and without the bright source to make details of the structure recovery more visible.
   }   
   \end{figure} 

\subsection{Recovery of parameters of Gaussian-shaped objects}
We initially examined simulated 12M array observations of a complex test image with groups of Gaussian blobs at a large range of angular scales (Fig. \ref{fig:defectimpact} a, b, c). We found that the results of the standard imaging process with 
CASA tclean\cite{casa2022}
are only weakly deteriorated by deviations of the BLD from the ideal {\it as long as a significant sensitivity to all relevant angular scales is maintained}. We determined that at least 77\% of the ideal sensitivity (corresponding to FF$=50$\%) needs to be retained at each angular scale range in order to keep image quality indistinguishable from that achieved with the ideal BLD.

\subsection{Simulation and analysis of stylised proto-planetary disks}
In order to study a more complex target shape than a Gaussian, we created
test images for each standard ALMA C43 containing concentric rings with dimensions 
ranging from close to the configuration's angular resolution 
to its maximum recoverable scale as given by the ALMA THB \cite{cortes2024}.
This setup is meant to resemble a stylised "proto-planetary disk", one of the typical targets of ALMA.
To challenge the deconvolution, the setup was studied with and without a bright compact source close to the edge of the outermost ring (Fig. \ref{fig:defectimpact} d, e, f).
   
Varying the BLD of the simulated array, we found again that the imaging
process is very robust to BLD defects until they begin to become larger
than 40\% over a range of ca. 20\% of the overall BL range.
But only when defects of 80\% (i.e. FF=20\%) are introduced in this way, do the 
corresponding deviations of the measured image parameters (like ring gap size, 
over-all flux etc.) become severe. This argues for a general lower limit on the FF of 25\%
corresponding to placing the upper limit on the acceptable noise RMS in a given uv coverage bin to {\it twice} the expected value.

Here and also in the general measurement of the MRS
for the standard ALMA configurations, we find that it is important to monitor
the homogeneity of the azimuthal coverage especially for the shortest baselines.
Generally, missing azimuthal coverage in a given BL range can cause very misleading
image artifacts if the observed object is bright at the
corresponding angular scales. The impact on general image fidelity is particularly
large when bright extended objects are observed with incomplete (but not absent)
uv coverage at the short baselines.

\subsection{Full astrophysical modeling of a real protoplanetary disk observation}
\begin{figure} [ht]
   \begin{center}
   \begin{tabular}{ccc} 
   \includegraphics[height=5cm]{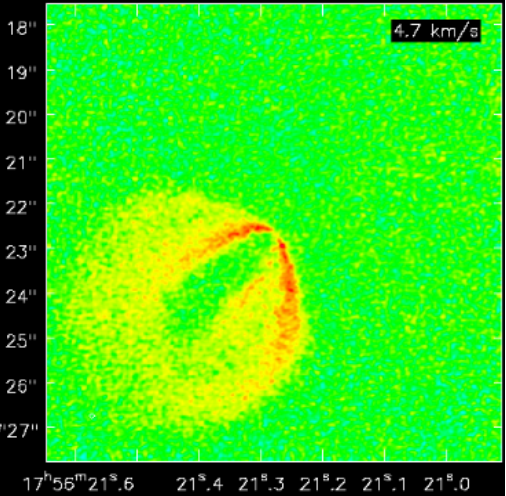} &
   \includegraphics[height=5cm]{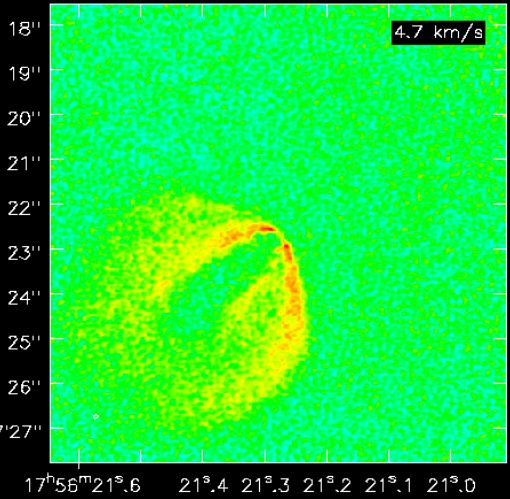} &
   \includegraphics[height=5cm]{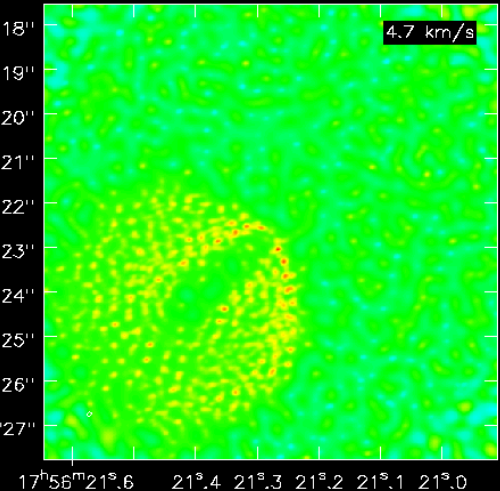} \\
   \end{tabular}
   \end{center}
   \caption[] { \label{fig:paneque}
   Narrow bandwidth (single-channel) image of the emission from a protoplanetary disk
   with different modifications to the BLD of the dataset before imaging:
   (left) ideal BLD, (middle) redistribution of sensitivity to remove 60\% of the visibilities from the lowest 10\% of the BLD, (right) redistribution of sensitivity  to remove 80\%  from the lowest 10\% of the BLD. All three datasets have the same 
   sensitivity when integrating over all angular scales.
   }   
   \end{figure}

\begin{figure} [ht]
   \begin{center}
   \includegraphics[height=5cm]{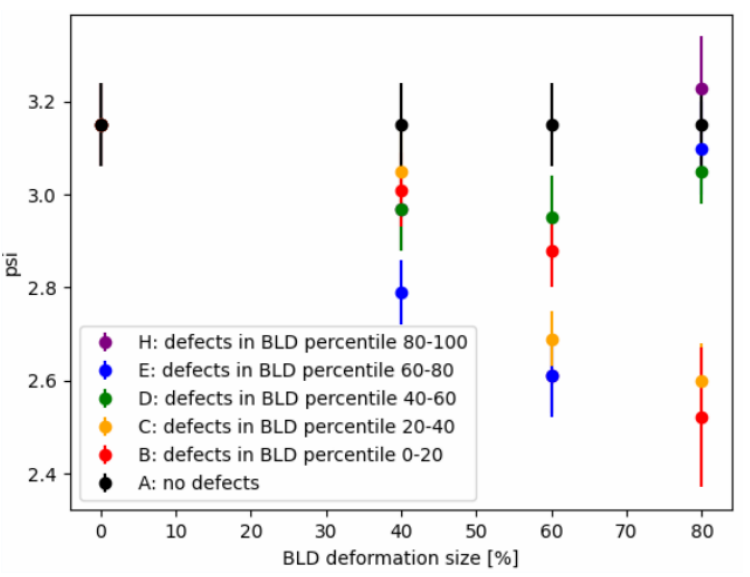} 
   \includegraphics[height=5cm]{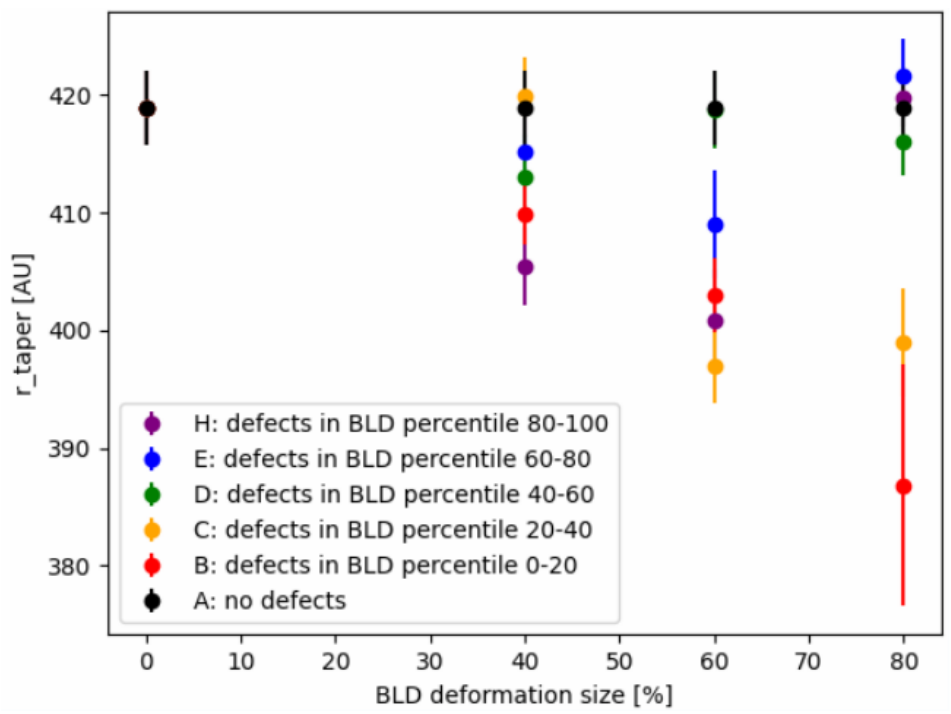} 
   \end{center}
   \caption[] { \label{fig:paneque2}
   Fit results on the different imagecubes from a CO line observation
   of a protoplanetary disk obtained with the ideal BLD (case A, black) and 
   deformed BLD (cases B, C, D, E, H, other colours) plotted as a function of the defect size. {\bf Left}: fit result for the parameter $\psi$. {\bf Right}:
   fit result for the parameter $r_{taper}$. The two parameters control the
   exponential decay of the altitude of the emitting surface of the protoplanetary
   disk over the plane of the disk as a function of radial distance from the
   disk center. The error bars shown represent the statistical errors provided
   be the fitting routine. As one can see, from a defect size of ca. 50\% onwards,
   the systematic errors caused by the BLD deformation dominate over the statistical errors.
   }   
   \end{figure}

We used code provided by T. Paneque-Carre\~{n}o \cite{paneque2023} to fit a well-tested model of protoplanetary disk structure to an ALMA dataset following a published procedure. By again varying 
the BLD of the input dataset while keeping the overall sensitivity constant,
we test how the fit results for the four model parameters react
to the different BLD defects. Fig. \ref{fig:paneque} shows examples
of the changes which the defects cause in the image cube,
while Fig. \ref{fig:paneque2} shows examples of the impact of these changes on
the model parameters.

Here we find that defect sizes of ca. 50\% are the tolerable limit. At this point, the parameter variations are as large as the statistical errors given by the fit algorithm. 
At a defect size of 80\%, the change caused by
the introduction of the BLD defect, in particular in the lower half of the BL range, can be several times larger than the nominal statistical error and thus inacceptable. The instrument is then simply no longer sensitive to the angular scales which dominate the emission from the protoplanetary disk.
This again argues for implementing at the very least a lower FF limit of 25\%.
Each BL range (averaged over azimuth) should contain at least 50\% of the expected visibility seconds,
i.e. have FF $\ge$~50\%.

\subsection{Other considerations and conclusions}
The deconvolution process using the CLEAN algorithm\cite{casa2022} is able to repair
shortcomings of the BLD, i.e. shortcomings of the PSF, sufficiently well unless
the angular scale and orientation in question is sampled with significantly lower sensitivity than 
the rest of the BLD. 

The exact limit to the tolerable BLD defects may depend on the 
science goal, but our results show consistently that
the deformation of the BLD away from the ideal shape by ca. 50\% in 
a range with a width of 20\% of the total baseline 
length range\footnote{which corresponds to a loss of roughly 23\% in sensitivity in that range and an equivalent gain in the remaining ranges since we are keeping the overall sensitivity constant}
is the point beyond which the effects on the scientific results of the
observation become significant and need to be included as additional
systematic errors. 

Our results show furthermore that a BLD underachievement by 75\% (50\% in sensitivity)
in a given angular scale range seems to be the maximum acceptable defect 
for an observation of an object with emission in that angular scale range. 
Observations with defects beyond this size
need special analysis techniques and can no longer be imaged in the conventional way.

Further arguments for keeping the BLD close to the ideal for a Gaussian PSF are the following:
\begin{enumerate}
\item A Gaussian PSF minimizes the noise-correlation in an interferometric image\cite{Tsukui2023}.
\item A Gaussian PSF minimizes the systematic errors on flux recovery\cite{Radcliffe2023}.
\item A Gaussian PSF minimizes the computing time needed for deconvolution since the flux to be redistributed in the image by
the deconvolution process is minimal. An observation
performed with a perfectly Gaussian beam would require no deconvolution 
at all\cite{boone2013}. 
\end{enumerate}

\section{Computing the expected BLD}
The ideal shape of the BLD can be derived from conditions on the shape of the PSF
of the interferometer and on the sensitivity at large angular scales. These conditions concern (a) the FWHM of a Gaussian fit to the PSF (corresponding to the achieved AR or smallest resolved angular scale), (b) the LAS (Largest Angular Scale of the science
target) or the MRS of the observation, and (c) the requirement of a general Gaussian shape of the PSF with minimal negative bowls. 

We have studied two methods to compute a BLD from the AR and MRS and find
that our "analytical" method is the preferred one\cite{petry2020} since it agrees in
most BL ranges with the present design of the ALMA configurations and is optimised for a 
Gaussian PSF. The "analytical" method computes the 2D BLD as a 2D Gaussian with an inner taper and then derives the 1D BLD. 

The MRS which the "analytical" BLD can achieve is, however, limited to ca. $8\times$AR. As mentioned above, we therefore need to introduce
a {\it modified analytical expectation} which can achieve the MRS values typically
requested from ALMA. This expectation is closer to the shape of the C43 configurations
because the design of these configurations followed the same reasoning.
It differs from the "analytical" BLD only in the lowest 10\% of the BL range
and at intermediate BLs near the peak of the BLD.

An alternative method for computing an expected BLD is what we call the "filled dish" (FD or 
"simulation") approach\cite{petry2020}. It fills a 2D aperture of a diameter corresponding to the AR
with randomly placed antenna positions imposing a minimum distance between all antennas corresponding to the MRS. We do not recommend to use this method for the ALMA 12M array but we intend
to study it further since it seems to have potential for BLD assessment in observations with
data combination between different 12M arrays and also for the ALMA 7M array.
The difference in BLD shape w.r.t. our first ("analytical") method consists mainly in a stronger emphasis on longer baselines with a cutoff at the BL which corresponds to the AR.
Furthermore, the FD method permits to extrapolate in a simple way to zero BL and may thus be
helpful in judging the BLDs of combined 12M array, 7M array, and single-dish (non-interferometric) datasets when pseudo-visibilities are introduced for the single-dish data.
Filled-dish array configurations have been studied in the past\cite{woody2001b} and found to perform well. 

The main reason why we recommend the "analytical" method over the FD method is that a pure FD BLD cuts off abruptly at a certain maximum BL and thus
does not optimize the side-lobe level, in disagreement with the present ALMA 12M configuration design. 

\section{Properties of the BLDs of real ALMA observations, MRS measurement, and the TM1+TM2 scheme}
\subsection{Underexposure at intermediate baseline lengths in the ALMA data} 

From our study of nearly 500 Cycle 6 and 7 and ca. 250 Cycle 9 12M MOUSs, we find that the ALMA 12M array observations show in general enough sensitivity at large and small angular scales but fall short of the requirements for an ideal PSF at {\it intermediate} scales. This is true regardless of the exposure time, the target elevation, and the number of antennas used. This is just a confirmation of what we have
seen in simulations in the nominal C43s, although there are significant differences among the configurations, with the compact ones being closer to the ideal\footnote{``ideal'' here refers to our recommended ``analytical'' method of computing the BLD expectation} and the more extended configurations deviating more significantly from it. As mentioned above, this is a consequence of the intention of the designers of the C43s to keep some large scale coverage while increasing sensitivity to shorter scales, all with the same total number of antennas.
Furthermore, some compromises with antenna relocation management were necessary.

\subsection{Exposure at the shortest baselines, MRS measurement, and implications for TM2} 
\label{sec:overexpshort}

The C43 design
results in an overexposure of most MOUSs at the shortest baselines compared to our
"analytical" expectation of the BLD shape. This means that (a) on the MOUS level, the PSF shape of this data is sub-optimal and (b) on the GOUS level, i.e. when combining several 12M array observations from
different configurations to achieve large angular scale coverage, at least the TM1+TM2 array combination groups have an even less
optimal PSF because here a TM2 (compact-configuration) observation is added to a TM1 (extended-configuration) observation to match a large requested MRS, which mostly increases
the sensitivity at short baselines where the deviation from an ideal BLD shape is already large.

We looked further into the question of scheduling and assessing
the uv coverage at short baselines by first developing methods to define and measure the 
MRS in a more stringent way.

\subsubsection{MRS and AR measurement via the "Flat sensitivity" plot}
Following a suggestion by Ryan Loomis (NRAO), we developed a new tool to compare the achieved sensitivity in a given BL range to an ideal, constant sensitivity over the entire observed angular scale range. This approach is related to methods developed for the analysis of the CMB\cite{2002MNRAS.334..569H}.
Based on information-theoretical considerations, the ideal of a sensitivity independent of angular scale is what an observer may naively assume when looking at an interferometric image. 

  \begin{figure} [ht]
   \begin{center}
   \includegraphics[height=5.8cm]{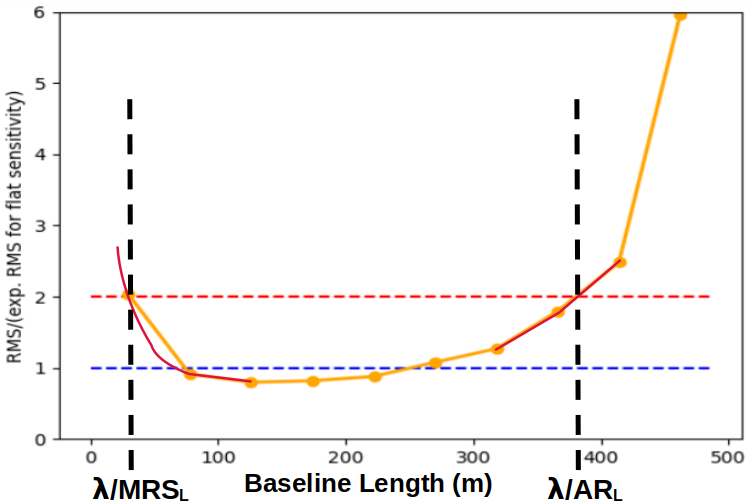}\\
   \end{center}
   \caption[] { \label{fig:flatsensplot} 
   Example of a "flat sensitivity" plot for a Cycle 9 observation of 6~min on-source time with
   the ALMA 12M array with 46 antennas in an intermediate configuration.
   The red curves indicate parabolic fits to the three points closest
   to the intersections with the red line of constant value 2.0 .
   The baseline lenghts corresponding to these intersections are
   used to define the values of MRS$_L$ and AR$_L$. See text.
   }   
   \end{figure}

The expected level of sensitivity in each BL bin of the BLD is determined such that 
the sensitivity is constant over the entire accessible angular scale range
in bins of equal width {\it in angular scale} and the overall sensitivity is
equal to the observed one.
Plotting the ratio of the actually achieved and the thus expected noise RMS 
in each BL bin versus BL gives a typically U-shaped graph indicating that in the middle
of the observed BL range, the achieved noise RMS is better than expected, and at the 
lower and upper end of the BL range the noise RMS diverges. We call this plot
the "flat sensitivity" plot. Fig. \ref{fig:flatsensplot} shows an example.

This plot permits to define the maximum recoverable scale MRS$_L$ as the angular scale corresponding
to the shortest BL where the achieved noise RMS is still smaller or equal to the 
max. acceptable RMS. We choose the max. acceptable noise RMS as $2 \times $ the expected RMS based on our results in section \ref{defectimpact}. In our plot of the noise RMS ratio this means
that the MRS$_L$ corresponds to the smallest BL where the plot reaches a value $\leq 2$. 
Similarly, also an AR can be defined as the angular scale corresponding to the {\it largest}
BL where the RMS is still $\leq 2 \times$ the expectation. We call this the AR$_L$.

Studying real data from Cycle 9 and simulations of all C43s, 
we find that AR$_L$ is well 
correlated with and on average equal to the standard measurements of AR via 
PSF analysis during image deconvolution.

The agreement of the MRS$_L$ with the MRS values derived from the L05 (as described
in the THB) is equally good both when measured in simulations and in real Cycle 9 data.

\subsubsection{Measurement of the MRS via imaging and flux recovery}
The MRS values in the ALMA THB were obtained by analysing the Fourier transforms of
uniformly filled disks and determining at which BL they drop below 10\% of the
theoretical maximum amplitude, the total flux of the disk,
i.e. by identifying the approximate position of the first minimum in the {\tt sync}-function-like
Fourier transform of the uniform disk\cite{remijan2019}.
This method is not intuitive for general ALMA users who think of their
observation targets in terms of all angular scales {\it combined}.

To provide an alternative measurement, we set up snapshot (3~min per field, 60~min total) simulated small mosaic observations 
of (a) Gaussian objects and (b) uniform disks of different sizes.
We vary the FWHM of the Gaussian (or the diameter of the disk respectively)
from a few times the THB AR value of the configuration to several times the THB MRS value.
We then image the resulting simulated visibilities with standard CASA "tclean" procedures\cite{casa2022} and measure the total flux of the object image. Finally, we  measure the flux recovery ratio as a function of object size and define
the MRS as the FWHM of the input Gaussian (or as the diameter of the disk) for
which the flux recovery drops below a certain limit, e.g. 50\%.
We find that this alternative way of measuring the MRS results in 
systematically larger values than those given by the THB. In particular for the disks, 
we obtain MRS values which come close to the theoretical limit of the MRS which is
$\lambda/\mathrm{BL}_{min}$.

For the Gaussians, we obtain MRS values which are closer to but still mostly above 
our MRS measurements from the "flat sensitivity" plots (MRS$_L$), which are 
themselves consistent with the MRS values derived from the L05 following the THB. 

In order to investigate the dependence of the MRS on noise level, we also study a low SNR case for the Gaussians which demonstrates that there is a significant
deterioration of MRS value when going from a dynamical-range-limited case
to a peak SNR of only 20.

All our MRS values are conservative since they are based on simulations at the
elevations of ca. 85$^\circ$. Observations at lower elevations will have at least 
partially foreshortened
baselines due to the projection effect and so the shortest baselines will be
up to a factor $\sin(\mathrm{elevation})$ shorter than their nominal length, but never
shorter than the dish diameter (because of shadowing).
This will correspondingly raise the MRS. The exact value will depend on
the hour-angle distribution and the Declination of the object.

On average, our MRS derived from the flux recovery
in high-SNR simulations of Gaussians is a factor 1.5 larger 
than the THB reference value, while the value derived from low-SNR simulations
is only a factor 1.36 larger.

In ALMA Cycle 9 data, we find that 97\% of the studied representative sample
of single-MOUS observations have achieved the requested LAS.

\subsection{General overexposure} 
In general, the sensitivity reached in ALMA observations up to Cycle 9 was often significantly 
better than requested. So methods to improve the BLD shape during offline analysis could be applied
without the overall sensitivity dropping below the requested value.
We have produced prototype code to demonstrate the adjustment of the BLD by re-weighting
but have to leave a detailed examination of the trade-off between improvement in PSF shape
and loss in sensitivity to the future.

In cases of bright objects, this approach looked particularly promising for
7M data which can have bad PSF shapes due to the 7M array's small number of baselines.
But first tests of the concept did not give good results. The 7M array is lacking 
baselines of 40~m - 70~m length to obtain a reasonably Gaussian PSF. 
For combination with 12M array data, however, this shortcoming of the stand-alone PSF
of the 7M array is irrelevant.

\subsection{Short observations} 
In the analysed sample of almost 500 Cycle 6/7 MOUSs form phase 1, we find that for Set~1, 71\% of all MOUSs consist of one execution. The on-source time is less than 5 minutes for 7\% of the sample and less than 15 minutes for 45\% of the sample. Only 21\% of all MOUSs have on-source times longer than 1 hour. Also because of these facts, it is important to specially consider short observations when investigating the properties of the ALMA 7M array and the 12M array in all but its most compact configurations. The 7M array and the extended 12M array have too inhomogenous uv coverage for snapshots with high image fidelity. 
A remedy would be to schedule two snapshot EBs with very different starting hour-angle.

\subsection{Observations with no explicit LAS request}
Essentially all observations will profit from a good PSF, and so
BLD assessment should be carried out on all of them.
The computational effort for BLD assessment is small and even if the PI of
a particular observation is not interested in high image fidelity, monitoring
the BLD of all observations will further help to understand operational constraints. 

  \begin{figure} [t!]
   \begin{center}
   \includegraphics[height=9cm]{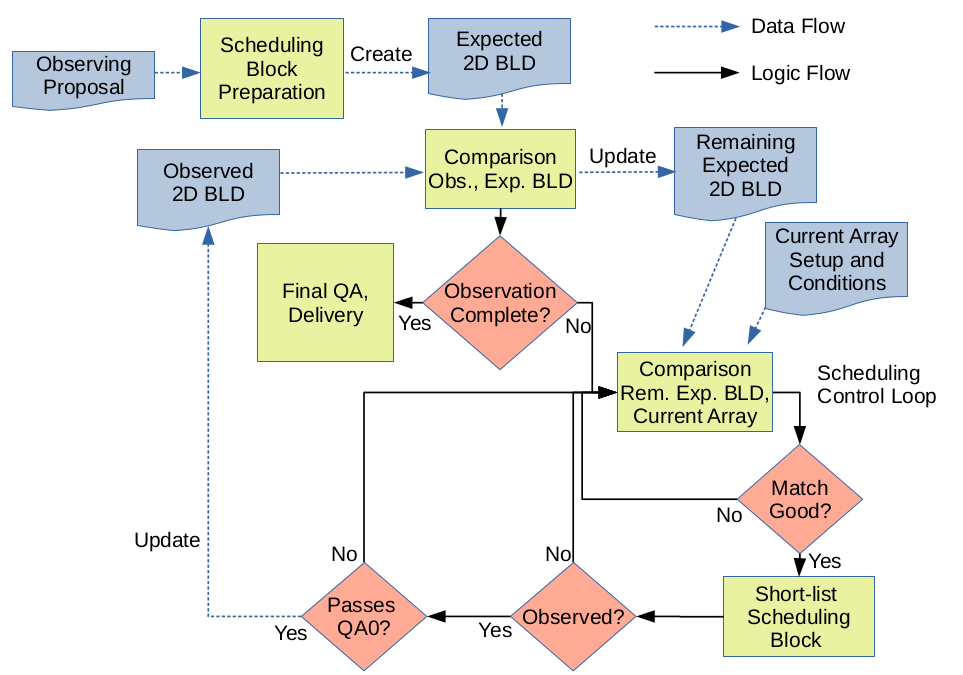}
   \end{center}
   \caption[] { \label{fig:controlloop} 
    Outline of our proposed new observing control loop with uv coverage tracking
    via the 2D Baseline Length Distribution (BLD). A scheduling block
    is preferentially observed when its remaining expected 2D BLD is a 
    good match with the properties of the current array and the general
    observing conditions.
   }   
   \end{figure} 
   
\section{Update to quality assurance}
We are testing an update to the uv coverage assessment in ALMA QA along the following lines: 
Based on our analytical expectation, the 4$\times$10 FF matrix should be computed for PSF assessment. 
Achievement of the AR and overall good PSF Gaussianity is verified by imposing the condition FF$>0.5$
on all ten BL bins (averaged over azimuth) and on all four azimuthal bins (averaged over BL),
i.e. each annulus and each sector in the uv plane should achieve at least 77\% of the
expected sensitivity.

The necessary low ellipticity of the synthesized beam in all BL ranges is enforced by requiring the FF for each of the ten BL ranges to be approximately constant over the four orientation bins. This homogeneity of 
the uv coverage in azimuth is measured separately in each BL range by fitting a constant function to the four FF values of the BL range and requiring the $\chi^2$ of the fit to be $\leq$2. 
Furthermore, no less than 90\% of the elements in the FF matrix must achieve
a minimum value of $0.25$, i.e. with few exceptions tolerated, each 2D bin in the uv plane should achieve at least 50\% of the expected sensitivity.

The achievement of the requested LAS is verified by comparing it to the MRS$_L$ value 
determined from the observed BLD using our "flat sensitivity" method. 

If the above criteria are not fullfilled by an MOUS, a more thorough inspection
is warranted. 

In a first run on ca. 250 MOUSs from ALMA Cycle 9 12M observations, 
ca. 50\% of the MOUSs did indeed pass this test, 40\% showed only minor defects (individual
FF matrix bins below 0.25 or individual BL bins being inhomogenous in azimuth)
while the remaining 10\% had defects like whole sectors or whole annuli having
a FF$<0.5$. In other words, 90\% of the ALMA 12M data had a uv coverage with minor or no defects. 
These numbers show that we are close to the right parameter range
for creating QA criteria which ALMA can fulfill but which are still sufficiently
ambitious. 

\section{Implications for scheduling}
The 7M array and the more extended 12M array configurations 
need Earth-rotation in order to achieve good coverage of all baseline orientations.
Our proposed 2D QA procedure (see above) will detect cases where additional observations
at different starting hour-angles (HA) are are needed. 

Scheduling multi-EB observations will profit from an intermediate uv coverage assessment when an MOUS cannot be completed in one session. We recommend to inspect the so-far achieved HA coverage 
during the initial per-EB quality assessment (QA0) and 
schedule the remaining EBs to extend the HA coverage as much as possible with
the remaining EBs.

In the ideal case, scheduling should keep track of the 2D BLD achieved so far in order to 
determine which array setup and which starting HA
is best to complete the MOUS and achieve a near-ideal BLD. In Fig. \ref{fig:controlloop} we
sketch how the observing control loop could be set up.

As part of this refactoring, the overall "execution fraction" scheme, which is presently in use
in ALMA scheduling and QA, would be replaced by one using our 4$\times$10 EF matrix.

\section{Software developed by this study}
We have developed a prototype for an ALMA uv coverage assessment
tool named "assess\_ms" and a number of Python modules for generating BLDs, operating on them, and analysing their properties. After "assess\_ms" has been introduced in ALMA science operations
and stood the test of a complete ALMA observing cycle, we are planning to make it publicly available 
for general ALMA users, in particular in the context of data combination.

  \begin{figure} [hbt]
   \begin{center}
   \begin{tabular}{ccc} 
   \includegraphics[height=4cm]{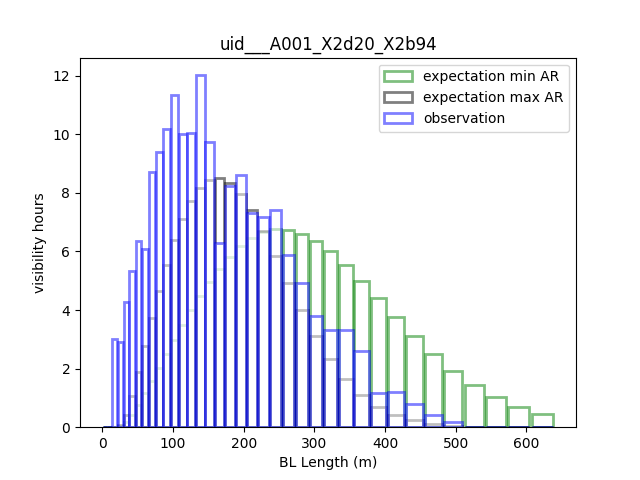} &
     \includegraphics[height=4cm]{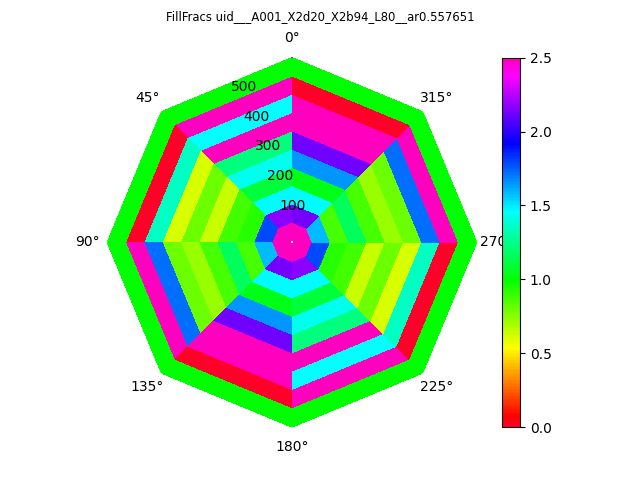} &
       \includegraphics[height=4cm]{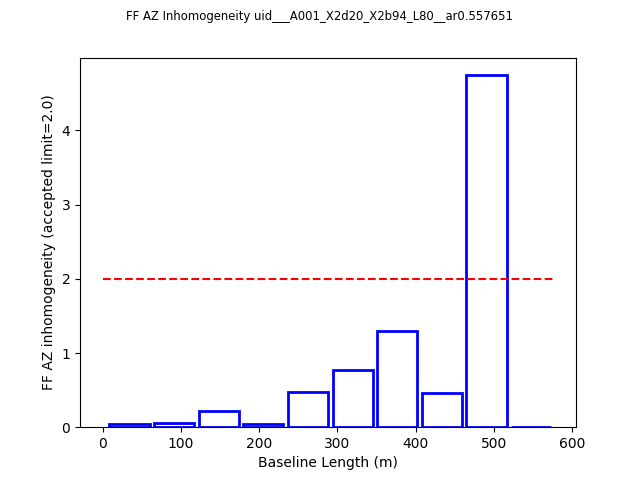} \\
         \includegraphics[height=4cm]{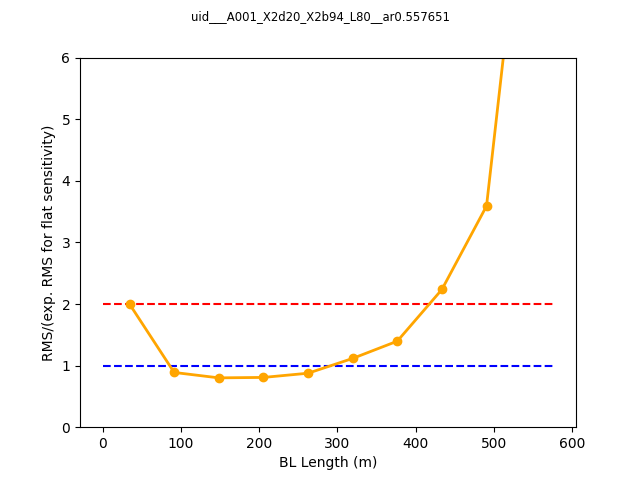} &
           \includegraphics[height=4cm]{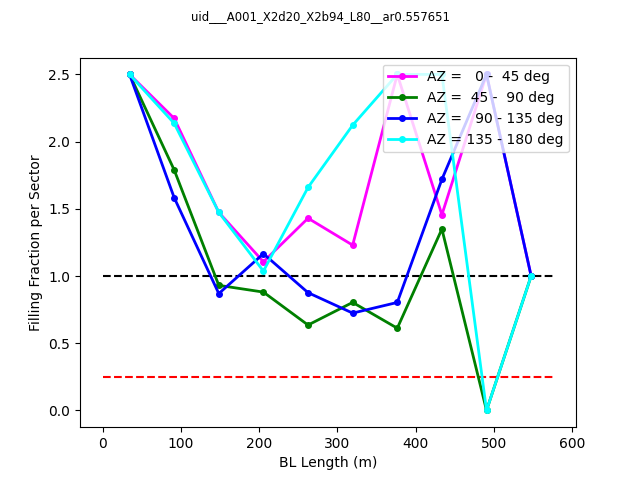} &
             \includegraphics[height=4cm]{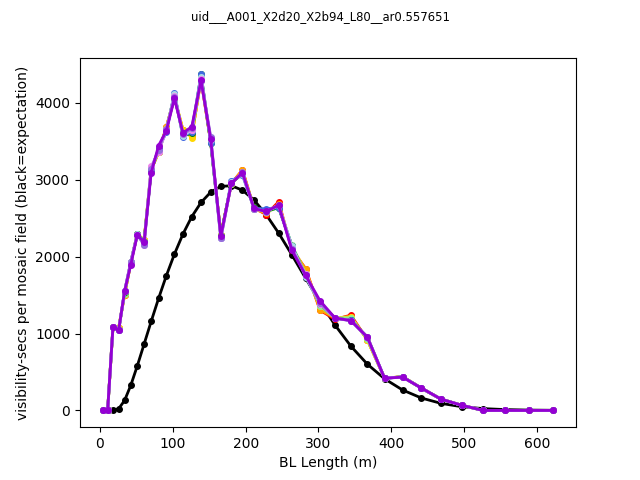} \\
   \end{tabular}
   \end{center}
   \caption[] { \label{fig:diagplots} 
      Example of the set of diagnostic plots for uv coverage assessment produced by our analysis tool "assess\_ms" for an ALMA observation
      (a mosaic) with some underexposure at long baselines in two sectors but otherwise good uv coverage:
      From left to right, first row: comparison of observed BLD with the two expectations for minimal and maximal acceptable AR,
      2D Filling Fraction plot, azimuthal inhomogeneity vs. BL (for the BL bin around 500~m, the
      inhomogeneity measure is above the threshold of 2.0). - Second row: "flat sensitivity" plot (see text), FF vs. BL for 
      each of the four sectors, BLDs for each of the mosaic fields overplotted (with the expected 
      shape for best PSF in black).
   }   
   \end{figure} 


\acknowledgments 
We would like to thank Carlos de Breuck (ESO) for his support as the contact person for this ESO ALMA internal development study.

\bibliography{thebibliog} 
\bibliographystyle{spiebib} 

\end{document}